\documentclass[nonacm, sigconf, anonymous = false]{acmart}

\usepackage{kantlipsum}
\usepackage{colortbl}
\usepackage{multicol}
\usepackage{graphicx}
\usepackage{caption,subcaption, ragged2e}
\usepackage{amsmath}
\usepackage{comment}
\usepackage{xspace}

\usepackage[british]{babel}
\usepackage{hhline}
\usepackage{multirow}
\usepackage{booktabs}
\usepackage{multirow}
\usepackage{siunitx}

\usepackage{array}
\usepackage{makecell}
\usepackage{graphicx}
\usepackage{multirow}
\usepackage{colortbl}
\usepackage{color}
\usepackage{tabularray}
\usepackage{tabularx}

\usepackage{pifont}

\usepackage{longtable}

\usepackage{xcolor,pifont}
\newcommand*\colourcheck[1]{%
  \expandafter\newcommand\csname #1check\endcsname{\textcolor{#1}{\ding{52}}}%
}
\colourcheck{blue}
\colourcheck{green}
\colourcheck{red}

\usepackage[marginparwidth=2cm]{}

\renewcommand\footnotetextcopyrightpermission[1]{} 

\newcommand{\x}{\mathbf{x}}
\newcommand{\w}{\mathbf{w}}
\newcommand{\btheta}{\ensuremath{\boldsymbol{\theta}}}
\newcommand{\bTheta}{\ensuremath{\boldsymbol{\Theta}}}
\newcommand{\bbR}{\ensuremath{\mathbb{R}}}

\newcommand{\calD}{\ensuremath{\mathcal{D}}}
\newcommand{\calL}{\ensuremath{\mathcal{L}}}

\makeatletter
\renewcommand\@formatdoi[1]{\ignorespaces}
\makeatother

\AtBeginDocument{%
  }

\setcopyright{none}
\copyrightyear{2018}
\acmYear{2018}
\acmDOI{XXXXXXX.XXXXXXX}

\begin{document}
\newcommand{\Name}{COMNETS}
\title{COMNETS: \underline{CO}st-sensitive decision trees approach to throughput optimization for \underline{M}ulti-radio IoT \underline{NET}work\underline{S}}


 \author{Jothi Prasanna Shanmuga Sundaram, Magzhan Gabidolla, Miguel {\'A}.\ Carreira-Perpi{\~n}{\'a}n, and \newline Alberto E. Cerpa}
 \email{{jshanmugasundaram,mgabidolla, mcarreira-perpinan, acerpa}@ucmerced.edu }
 \affiliation{Department of Computer Science and Engineering, University of California, Merced \country{USA}}

\begin{abstract}

Mesoscale IoT applications, such as P2P energy trade and real-time industrial control systems, demand high throughput and low latency, with a secondary emphasis on energy efficiency as they rely on grid power or large-capacity batteries. MARS, a multi-radio architecture, leverages ML to instantaneously select the optimal radio for transmission, outperforming the single-radio systems. However, MARS encounters a significant issue with cost sensitivity, where high-cost errors account for 40\% throughput loss. Current cost-sensitive ML algorithms assign a misclassification cost for each class, but not for each data sample. In MARS, each data sample has different costs, making it tedious to employ existing cost-sensitive ML algorithms. First, we address this issue by developing \Name, an ML-based radio selector using oblique trees optimized by Tree Alternating Optimization (TAO). TAO incorporates sample-specific misclassification costs to avert high-cost errors, and achieves a 50\% reduction in the decision tree size, making it more suitable for resource-constrained IoT devices. Second, we prove the stability property of TAO and leverage it to understand the critical factors affecting the radio-selection problem. Finally, our real-world evaluation of \Name\xspace at two different locations shows an average throughput gain of 20.83\%, 17.39\% than MARS. 
\end{abstract}

\maketitle
\pagestyle{plain} 

\vspace{-0.1in}
\section{Introduction}

The Internet of Things (IoT) is rapidly utilized in day-to-day lives for asset monitoring, home automation, and Industrial automation as it reduces the human effort involved. Conventional IoT applications were utilized in small-scale deployments like home automation, that span up to 100m. These applications utilize short-range radios like Zigbee and Bluetooth Low-Energy (BLE). Due to recent advancements in the IoT radio technology, long-range radios like WiFi-HaLow, LoRa, Sig-Fox, and NB-IoT were developed \cite{sundaram2019survey} for large-scale applications, like Microsoft Farmbeats \cite{vasisht2017farmbeats}, ranging between 1-5Km. The applications that range between 100-1000m, like Industrial automation\cite{fu2024comprehensive}, P2P energy-trade in smart meters \cite{mitsubishielectricPowerSystems, mitsubishielectricSmartMeter}, are termed as the mesoscale IoT applications.   

\textbf{Mesoscale IoT applications} are identified as the new and upcoming category of IoT applications \cite{sundaram2024mars}. Unlike the traditional IoT applications that have energy constraints, the mesoscale applications are either grid-powered \cite{nagai2024improve, yang2024rateless, tan2018resilience} or have a large-battery reserve \cite{mitsubishielectricSmartMeter, mitsubishielectricLargecapacityBattery, mitsubishielectricPowerSystems}. These mesoscale applications range between 100-1000m, with no specialized radios developed to cater to these applications~\cite{sundaram2024mars}. To solve this issue, MARS \cite{sundaram2024mars} employs a multi-radio architecture comprising of multihop Zigbee and single hop LoRa radios with radio selection using axis-aligned trees optimized by the Tree Alternating Optimization (TAO) \cite{carreira2018alternating}. MARS identifies the \textit{gray-region}, 500-1200m from the gateway, where Zigbee and LoRa radios achieve competitive throughput. Due to the erratic link quality variations in the mesoscale environment, it is uncertain which radio will achieve better throughput at a given time instant. So, MARS formulated the radio selection problem as a classification problem with 0/1 loss, to maximize the throughput.           

\textbf{Classification problems in machine learning} focus on maximizing classification accuracy, assuming equal cost for all misclassification errors \cite{jan2012simple, lin2019advances, domingos1999metacost}. However, classification accuracy is not the only important factor in many real-world applications, and each misclassification error may incur different costs \cite{elkan2001foundations, kim2012classification}. 

For example, in brain tumor detection applications, wrongly identifying a healthy patient as a tumor patient has a different cost than identifying a tumor patient as a healthy patient. Churn modeling identifies the customers that are more likely to leave a service provider. Accurately identifying the churner is the task at hand but identifying the profitable churner and unprofitable churner is more important in this context \cite{glady2009modeling}. In marketing applications, wrongly identifying that a specific customer will not accept the offer when they are ready to accept the offer costs money while the vice-versa prediction error costs time \cite{zadrozny2003cost}. Finally, in intrusion detection applications, classifying malicious connections as benign has a different cost than the vice-versa classification.
\begin{table}[t]
  \centering
  \caption{Cost-sensitivity in radio selection}
  \vspace{-0.1in}
    \label{tab:cost_radio}
  \begin{tabular}{ccl}
    \toprule
                        & T1    & T2\\  
    \toprule
    Zigbee Throughput   & 7000        & 4600  \\
    LoRa Throughput     & 2000        & 4500  \\
    Throughput Loss     & 5000        & 100   \\
    Category            & High-cost   & Low-cost   \\
    \bottomrule
  \end{tabular}
  \vspace{-0.3in}
\end{table}

\textbf{Cost-sensitivity in radio selection.} Table. \ref{tab:cost_radio} shows the importance of cost in the radio selection problem. In the radio-selector problem, when one radio is achieving higher throughput than the other, the difference in throughput will not be the same for all the instances. For example, at T1,  let's assume that the throughput of the ZigBee radio is 7000~bps and the throughput of LoRa is 2000~bps. The throughput difference between these two radios at this time instance is 5000~bps. At T2, say the throughput of a Zigbee radio is 4600~bps and the throughput of a LoRa radio is 4500 bps. The throughput difference is only 100~bps. At T1, assuming that the radio selector algorithm misclassifies the low-throughput LoRa radio as the high-throughput radio costs 5000~bps, while a misclassification at T2 costs only 100~bps. Misclassification at time T2 is tolerable since there is little throughput gain, while the misclassification at time T1 is intolerable for the multi-radio system. 

\textbf{Cost-sensitive ML algorithms} in the literature define a cost matrix associated with misclassifying each class. Here, the misclassification cost is always tied to a class instead of a specific data sample~\cite{elkan2001foundations}. In the radio selector problem, the cost varies for each data sample. Cost-sensitive optimization for each data sample is tedious (refer \S\ref{sec:background} for more details). An intuitive way is to utilize the cost in the loss function of the ML models. Our experiments with five-fold cross-validation show that this method gives a model that is not well generalized for unseen data (refer \S\ref{sec:ML_model}). 

First, we employed Oblique-trees optimized with TAO\cite{carreira2018alternating} algorithm to solve this problem. TAO\cite{carreira2018alternating} does this optimization gracefully to avoid high-cost errors as explained in \S\ref{sec:background}. To the best of our knowledge, we are the first to employ TAO\cite{carreira2018alternating} to optimize Oblique trees for this cost-based optimization. Compared to the other commonly used models like SVM, Logistic Regression and CART, TAO-Oblique trees achieve the best results. The TAO\cite{carreira2018alternating} optimization algorithm not only optimizes to avoid high-cost errors, but also makes the trees 50\% smaller compared to the other models. 

Second, we interpret the essence of the radio selection problem using the tree stability property of TAO\cite{carreira2018alternating} algorithm. TAO not only optimizes cost-sensitive oblique trees and makes them smaller, but it also provides stability to the tree when new data are added. In general, the decision trees (axis-aligned/Oblique) are easy to interpret. The reason they are not extensively used is that when new data are added, they induct a completely new tree with different rules \cite{turney1995bias, li2002instability}. TAO\cite{carreira2018alternating} optimization on Oblique trees overcomes this issue by providing stability to the tree, i.e. TAO optimization reduces this drastic change in the tree structure even when new data are added. This helps to understand the essence of the radio selection problem at hand. Since TAO-Oblique trees use a linear combination of multiple features, it is tedious to extract a meaningful reason from the interpretations. So, we use TAO\cite{carreira2018alternating} optimized sparse oblique trees to solve this issue. Sparse-oblique trees reduce the number of parameters used in each decision node, making it easier to extract a meaningful reason for the interpretations.

Finally, the TAO\cite{carreira2018alternating}-Oblique trees are converted into IF...ELSE statements for easier deployment in resource-constrained IoT devices. Our real-world, large-scale evaluation of \Name\xspace, powered by TAO\cite{carreira2018alternating}-Oblique tree, shows a throughput gain of 20.83\% \& 17.39\% over the state-of-the-art MARS, at locations A \& B respectively. 

In summary, the contributions of our work are:
\begin{itemize}
    \vspace{-0.15cm}
    \item We identified and showed the importance of cost-sensitivity in multi-radio networks through real-world experiments. 
    \item We developed TAO\cite{carreira2018alternating}-Oblique trees that optimize to avert high-cost classification errors. 
    \item We provided insights into the radio selection problem by exploiting the stability property of the TAO\cite{carreira2018alternating} algorithm.
\end{itemize}


\section{Related Work}
\label{sec:related_work}

Bahl et al. \cite{bahl2004reconsidering} showed that a multi-radio system is beneficial for wireless networks. A lot of multi-radio wireless systems \cite{ananthanarayanan2009blue, sur2017wifi} has been developed to optimize the performance of the networks like energy-efficiency~\cite{jin2011wizi, kusy2014radio, lymberopoulos2008towards} and routing management~\cite{draves2004routing}. Backpacking~\cite{al2011backpacking} was developed for high data rate sensor networks. Kusy et al.~\cite{kusy2014radio} developed a multi-radio architecture for WSN. They show that employing two multi-hop radios in the same node improves reliability with 3-33\% energy overhead. LoRaCP \cite{gu2019one} employs a ZigBee+LoRa multi-radio network for faster control packet transmissions in multi-hop WSN. Gummeson et al.~\cite{gummeson2009adaptive} optimize
energy consumption by employing a Reinforcement Learning (RL) based adaptive link layer to switch radios (CC2420+XE1205) based on channel dynamics. Using RL is detrimental because of: (i) Very high training data requirements and, (ii) inference latency\cite{sundaram2024mars}. Lymberopoulos et al.~\cite{lymberopoulos2008towards} switch radios (Zigbee+WiFi) with a threshold-based algorithm optimizing for energy efficiency.

While most of the above multi-radio systems developed for IoT networks optimize for energy efficiency over small-scale deployments, MARS \cite{sundaram2024mars} optimizes for throughput and latency on mesoscale applications. Initially, they identified the absence of a fully developed specialized radio for emerging mesoscale IoT applications. To close this gap, they first conducted a qualitative analysis on the suitability of all the available IoT radios for mesoscale application environments and identified that Zigbee2.4GHz and LoRa915MHz are the better candidates. Their analytic and experimental analysis with these two radios shows that Zigbee and LoRa achieve competitive throughput in the \textit{gray-region}, 500-1200m from the gateway~\cite{sundaram2024mars}. Since it is uncertain which radio will provide higher throughput at the time of transmission, they developed a Decision Tree (DT) model using axis-aligned trees to predict the high-throughput radio and further optimize this DT model with Tree-Alternating Optimization (TAO) algorithm\cite{carreira2018alternating}. This ML model needs the link quality of LoRa and the instantaneous path quality estimations of the Zigbee radios. The traditional path quality estimations of Zigbee are not instantaneous. Hence, they develop a DT-based instantaneous path quality estimation to provide instantaneous inputs for the ML model. On taking the instantaneous inputs, their ML model will output the high-throughput radio to transmit the scheduled packet.
\begin{figure}[t]
   \centering
   \includegraphics[scale=0.6]{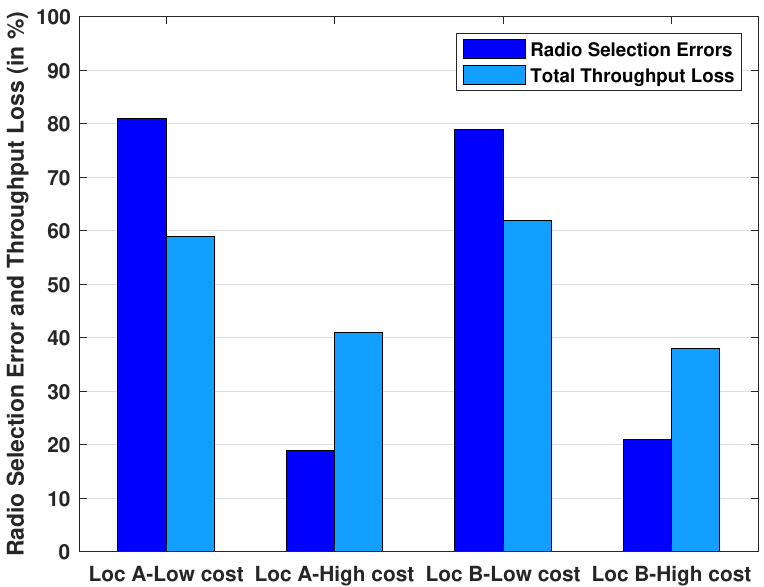}
        \vspace{-0.1in}
        \caption{Motivation for cost-sensitive learning}
        \label{fig:motivation}
\end{figure}
\vspace{-0.2in}
\section{Motivation}
\label{sec:motivation}
We quantified this cost sensitivity with traces obtained from real-world experiments. 
From these traces, we categorized the radio selection errors as high-cost and low-cost errors. Low-cost errors have a throughput difference between two radios less than 200 bps. High-cost errors have a throughput difference greater than 200 bps. 

Figure \ref{fig:motivation} shows our motivation to pursue cost-sensitive learning for this radio selector problem. In Location A, the low-cost errors amount to 81\% of the total radio selection errors costing 59\% of the total throughput loss. The high-cost errors amount to 19\% of the total radio selection errors costing 41\% of the total throughput loss. Even though the difference between the number of high-cost and low-cost errors is very high, the difference between the total throughput loss between these two categories of errors is very low. A similar trend is seen at Location B. If the decision trees are optimized to avoid high-cost errors, nearly 40\% of the total throughput loss can be avoided in the multi-radio system. 
\begin{table}[t]
\centering
\caption{Throughput gain and optimization margin}
\label{tab:margin}
\resizebox{\columnwidth}{!}{%
\begin{tabular}{|c|cc|}
\hline
\multirow{2}{*}{} & \multicolumn{2}{c|}{Average throughput gain (in \%)} \\ \cline{2-3} 
                  & \multicolumn{1}{c|}{Location A}     & Location B     \\ \hline
MARS              & \multicolumn{1}{c|}{49.79}          & 48.20           \\ \hline
PERFECT OPTIMIZER   & \multicolumn{1}{c|}{73.23}          & 69.50          \\ \hline
OPTIMIZATION MARGIN            & \multicolumn{1}{c|}{23.44}          & 21.30          \\ \hline
\end{tabular}%
}
\end{table}

The throughput gain and optimization margin are tabulated in Table \ref{tab:margin}. The average throughput gain of MARS in Location A and Location B is 49.79\% and 48.2\% respectively. Assuming that a perfect optimizer algorithm can avert all high-cost errors, the average throughput gain of the multi-radio system at Locations A \& B will be increased to 73.23\% and 69.50\% giving an optimization margin of 23.44\% and 21.30\% respectively. This motivates us to pursue the cost-sensitivity problem for multi-radio IoT networks.

\section{Optimizing cost-sensitive loss}
\label{sec:background}

From a machine learning perspective, our problem consists of a cost-sensitive binary classification problem, where the two classes are LoRa, ZigBee radios. At any point in time for a given state of the system, if the ML model makes the correct prediction, i.e. chooses the radio with the highest throughput, then there is no corresponding loss for this case. But if the model makes an incorrect prediction at a given time instance, then there is a corresponding loss, which is the opportunity cost of choosing the radio with better throughput, whose amount is instance/time-specific. In order to train an ML model that optimizes for the best possible throughput, we first need to define a desired objective (loss) function.

Let $\{\x_n, y_n, c_n\}_{n=1}^N$ be our training set of $N$ points, where $\x \in \bbR^D$ is a $D$-dimensional feature vector describing the current state of the system, $y \in \{0, 1\}$ is a class label indicating the radio with better throughput, and $c \in \bbR_{>0}$ is a cost (weight) of the instance obtained by the difference between two types of radio throughputs. Let $T(\x; \bTheta)\mathpunct{:}\ \bbR^D \to \{0,1\}$ be an oblique decision tree with learnable parameters $\bTheta$. Based on the aforementioned discussion on the importance of costs in our setting, we aim to optimize the loss function:
\begin{equation}
    \label{e:loss-func}
    \min_{\bTheta} L(\bTheta) = \sum_{n=1}^{N}{c_n \cdot L(y_n, T(\x_n; \bTheta))}
\end{equation}
where $L(\cdot,\cdot)$ is a 0/1 loss. With the inclusion of costs $\{c_n\}_{n=1}^N$, the objective function of \eqref{e:loss-func} puts more importance on instances where there is a large gap between the throughputs of the two radio types while misclassifying instances with small throughput differences is less important. It captures exactly what we are after in our radio selection problem: better overall throughput. However, the loss function $L(\cdot,\cdot)$ is not differentiable, and with another highly non-differentiable function as a decision tree $T(\x; \bTheta)$ makes the learning problem of \eqref{e:loss-func} much harder to optimize.

To optimize a cost-sensitive 0/1 misclassification loss over a decision tree $T(\x; \bTheta)$, we rely on the TAO algorithm. It is a general algorithmic framework that can optimize different types of loss functions over various tree-based methods, and we will apply it here to learn cost-sensitive oblique decision trees. At a higher level, the TAO algorithm operates very differently than traditional tree learning methods such as CART or C5.0. Instead of growing a tree greedily starting from a root node based on some impurity criteria, it takes an initial tree of some predetermined fixed structure and initial parameters (e.g.\ a complete tree of depth $\Delta$ or a greedily induced tree), and performs alternating optimization steps over each node parameters, and with each such step guaranteeing a monotonic decrease of an objective function. Conceptually, it operates similarly to how neural networks are learned: defining a model architecture (cf.\ tree structure) and initial parameters (cf.\ tree parameters), and optimizing them with gradient-based methods (cf.\ alternating optimization). Below we describe more formally the decision tree $T(\x; \bTheta)$, and provide the specific details of optimization steps in TAO for the cost-sensitive loss.
\begin{figure*}[t]
  \captionsetup[subfigure]{justification=Centering}
  \centering
  \begin{subfigure}{0.48\textwidth}
    \includegraphics[width=0.9\columnwidth, height = 3.5cm]{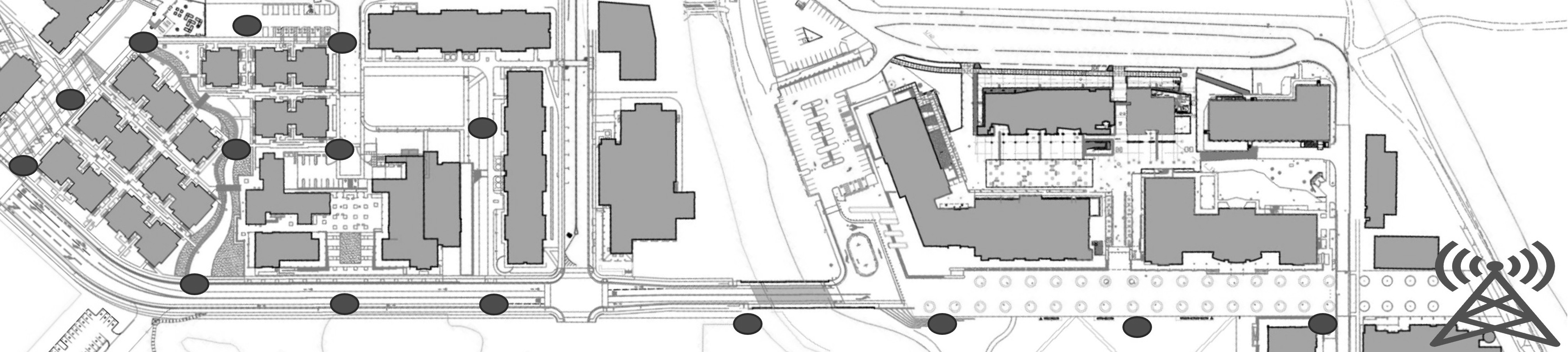}
    \caption{Location A - Mesh topology}
    \label{fig:LocA-mesh-topo}
  \end{subfigure}
  \begin{subfigure}{0.48\textwidth}
    \includegraphics[width=0.9\columnwidth, height = 3.5cm]{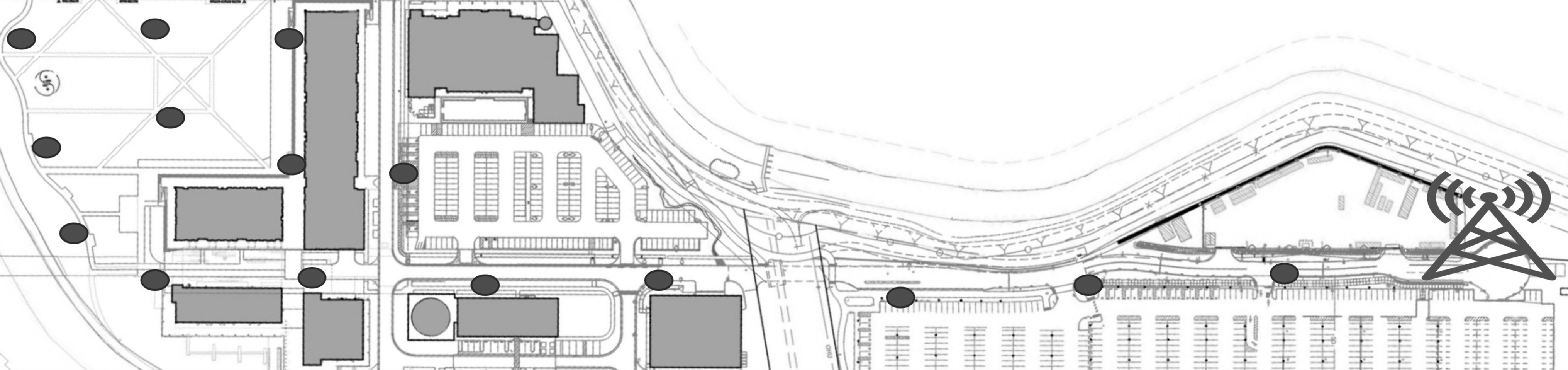}
    \caption{Location B - Mesh topology}
    \label{fig:LocB-mesh-topo}
  \end{subfigure}
  \caption{Mesh topology set up at locations A and B}
\end{figure*}
\begin{figure}[h]
   \centering
   \includegraphics[width=0.9\columnwidth, height=4.3cm]{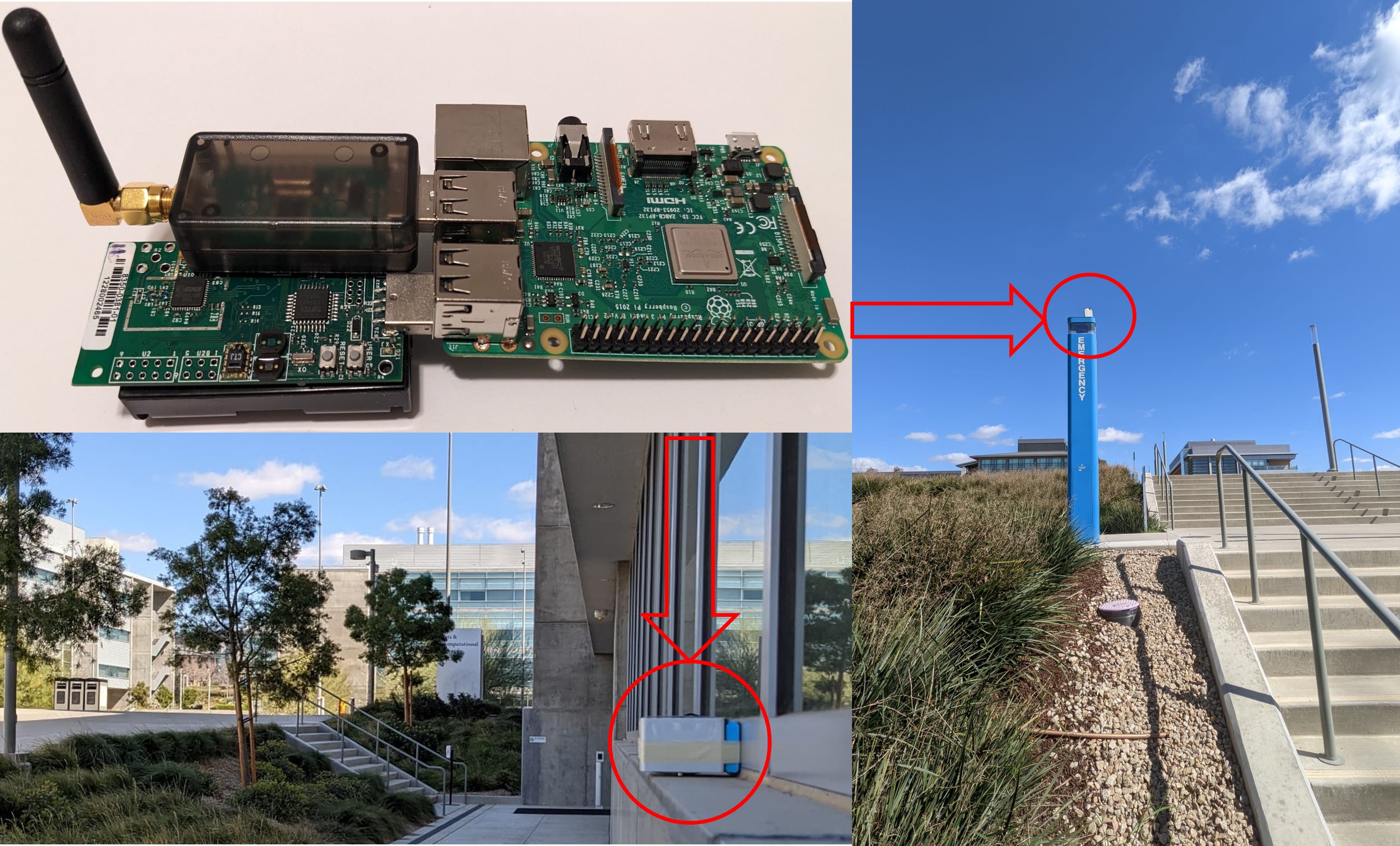}
        \caption{Multi-radio hardware}
        \label{fig:hardware}
\end{figure}
The predictive function $T(\x; \bTheta)$ of a decision tree works by routing an instance $\x$ through a root-to-leaf path where each decision node $i \in \calD$ ($\calD$ is the set decision node indices, $\calL$ is the set of leaf indices) in the path uses a decision function $g_i(\x;\btheta_i)\mathpunct{:}\ \bbR^D \to \{\text{left}_i, \text{right}_i\} \subset \calD \cup \calL$ to decide which child node the instance $\x$ will be sent next. In this paper, we consider oblique trees that use a linear model at decision nodes: $g_i(\x;\btheta_i) = \w_i^T \, \x + \w_{i0}$ with learnable parameters $\btheta_i = \{\w_i, \w_{i0}\}$. The actual prediction of the tree $T(\x; \bTheta)$ happens in the leaf, where it just outputs the leaf's class label i.e.\ the radio type. So a leaf $j \in \calL$ has only a single learnable parameter $\theta_j \in \{0, 1\}$. With learnable parameters explicitly defined, we now include regularization (penalty) terms on them in the objective function. The main purpose of this is to obtain models with better generalization performance as well as remaining small in the number of parameters and size. In particular, we include $\ell_1$ norm penalty on decision node hyperplanes. This helps to produce sparse hyperplanes, i.e.\ those with few nonzero weights, and if the regularization term is high enough, then the entire weight vector will turn to zero. A decision node with few nonzero weights are typically more interpretable, and the one that has entirely zero hyperplane sends all instances to only one child ($g_i(\x;\btheta_i) = \mathbf{0}^T \, \x + \w_{i0} = \w_{i0}$), whereby an entire other child subtree can be pruned. With regularization terms included, the final objective function that we optimize is the following:
\begin{equation}
    \label{e:obj-func}
    \min_{\bTheta} E(\bTheta) = \sum_{n=1}^{N}{c_n \cdot L(y_n, T(\x_n; \bTheta)) + \lambda \, \sum_{i \in \calD}{\lVert \w_i \rVert}_1}
\end{equation}

Given some tree structure and initial parameters, the TAO algorithm optimizes the above objective by monotonically decreasing it in each step. It is based on the following theorems:
    
\textbf{Seperability.} states that the objective function~\eqref{e:obj-func} separates over sets of any non-descendant nodes in the tree given the other fixed nodes. This stems from the fact that the tree makes hard decisions, and that the training points that reach any set of non-descendant nodes are disjoint. The implications of this theorem are that the non-descendant set of nodes can be independently optimized in parallel. The exact form of the node optimization depends on the node type (i.e.\ a decision node or a leaf) as explained below.

\textbf{Reduced problem over decision nodes.} Optimizing the top-level objective function~\eqref{e:obj-func} over the parameters of a decision node $i \in \calD$ reduces to a simpler problem of a weighted 0/1 loss binary classification that depends only on the training instances that reach this node $i$. This follows from the fact that the only thing a decision node can do is to send a point $\x_n$ either to the left or to the right child subtrees. If both child subtrees classify the point $\x_n$ correctly (or incorrectly), then it does not matter where to send it, and so those points can be discarded from the reduced problem. But if one child subtree classifies it correctly, while the other does not, then we want the decision function $g_i(\x_n; \btheta_i)$ to learn to send it to the ``correct'' child subtree. Repeating this observation for all training points in the reduced set, we can form a binary classification problem, where the ground truth binary labels correspond to the ``correct'' child for each point. And the problem is weighted because each training instance carries the original costs from the top-level objective function. There is also an $\ell_1$ penalty on decision node weights $\lVert \w_i \rVert_1$ that directly comes to the reduced problem from the top-level objective. Solving exactly a 0/1 misclassification loss over a linear model (and with $\ell_1$ penalty) is in general NP-hard problem, but we can approximate this reduced problem by a convex surrogate such as logistic regression. In our implementation, we use an $\ell_1$-regularized logistic regression solver inside the LIBLINEAR library \cite{liblinear}. To ensure a monotonic decrease of the objective, we can update the decision node parameters by the logistic regression solution only if it improves over the previous one in terms of the weighted 0/1 loss.

\textbf{Reduced problem over leaf nodes.} The top-level optimization problem~\eqref{e:obj-func} over a leaf node $j \in \calL$ is simply the same loss function but over a leaf parameter $\theta_j$ and over the set of training points that reach this leaf. With simple constant leaf nodes, the optimal solution is just a weighted majority class.

The theorems underlying the TAO algorithm tell us how individual nodes must be optimized but do not prescribe the order the nodes must be visited. We follow the reverse breadth-first-order approach similar to the original paper on TAO \cite{carreira2018alternating}: all nodes at the deepest level (from root) are optimized first, then their parents, and so on, until the root node. This is repeated multiple times (at most 20) until the objective function converges. As for the initial tree structure and parameters, we experiment with either a random complete tree of depth $\Delta$ or the one obtained from a CART algorithm and choose the best one (in terms of validation error).  

\vspace{-0.3cm}
\section{COMNETS ML model}
\label{sec:ML_build}
In this section, we explain the multi-radio system we built, the topology of our deployments for data collection, and the process of building the machine learning models. 

\textit{\textbf{Multi-radio hardware}} is shown in Figure~\ref{fig:hardware}. This consists of a Raspberry Pi~3B hosting two radios, (i)USB-based TelosB~\cite{telosb} Zigbee mote and (ii) a USB-based LoStik LoRa~\cite{ronoth_llc} radio. This multi-radio node is powered by a portable external power bank. We protected this node using a PVC case during deployments. The Raspberry Pi 3B will send a command to both the TelosB~\cite{telosb} and LoStik~\cite{ronoth_llc} radios, once every 3 seconds, to transmit a 29-byte packet, destined to the gateway, concurrently. LoRa radios employ ALOHA MAC while Zigbee radios use CSMA MAC. Link-level acknowledgments are disabled in the TelosB~\cite{telosb} Zigbee motes. 
\begin{table*}[t]
\caption{Cost-weighted accuracies of different machine learning models and optimizations.}
\label{tab:prediction_models}
\centering
\resizebox{\textwidth}{!}{%
\begin{tabular}{|c|cccc|cccc|}
\hline
\multirow{2}{*}{\begin{tabular}[c]{@{}c@{}}ML \\ Models \& \\ Optimizations\end{tabular}} &
  \multicolumn{4}{c|}{Location A} &
  \multicolumn{4}{c|}{Location B} \\ \cline{2-9} 
 &
  \multicolumn{1}{c|}{\begin{tabular}[c]{@{}c@{}}Training CWA\\ (in \%)\end{tabular}} &
  \multicolumn{1}{c|}{\begin{tabular}[c]{@{}c@{}}Testing CWA\\ (in \%)\end{tabular}} &
  \multicolumn{1}{c|}{Depth} &
  No. of Leaves &
  \multicolumn{1}{c|}{\begin{tabular}[c]{@{}c@{}}Training CWA\\ (in \%)\end{tabular}} &
  \multicolumn{1}{c|}{\begin{tabular}[c]{@{}c@{}}Testing CWA\\ (in \%)\end{tabular}} &
  \multicolumn{1}{c|}{Depth} &
  No. of Leaves \\ \hline
SVM &
  \multicolumn{1}{c|}{92.48 $\pm$ 0.33} &
  \multicolumn{1}{c|}{92.66 $\pm$ 0.66} &
  \multicolumn{1}{c|}{-} &
  - &
  \multicolumn{1}{c|}{80.12 $\pm$ 0.54} &
  \multicolumn{1}{c|}{82.74 $\pm$ 0.80} &
  \multicolumn{1}{c|}{-} &
  - \\ \hline
LR &
  \multicolumn{1}{c|}{92.64 $\pm$ 0.35} &
  \multicolumn{1}{c|}{92.78 $\pm$ 0.44} &
  \multicolumn{1}{c|}{-} &
  - &
  \multicolumn{1}{c|}{80.94 $\pm$ 0.52} &
  \multicolumn{1}{c|}{80.40 $\pm$ 1.37} &
  \multicolumn{1}{c|}{-} &
  - \\ \hline
CART &
  \multicolumn{1}{c|}{\textit{\textbf{97.68 $\pm$ 0.42}}} &
  \multicolumn{1}{c|}{93.56 $\pm$ 1.11} &
  \multicolumn{1}{c|}{8.0 $\pm$ 0.6} &
  31.6 $\pm$ 5.7 &
  \multicolumn{1}{c|}{\textit{\textbf{92.25 $\pm$ 2.27}}} &
  \multicolumn{1}{c|}{85.88 $\pm$ 1.49} &
  \multicolumn{1}{c|}{8.8 $\pm$ 4.1} &
  33.8 $\pm$ 24.1 \\ \hline
TAO-Axis Aligned &
  \multicolumn{1}{c|}{96.36 $\pm$ 1.09} &
  \multicolumn{1}{c|}{94.67 $\pm$ 1.68} &
  \multicolumn{1}{c|}{7.2 $\pm$ 1.3} &
  25.2 $\pm$ 4.8 &
  \multicolumn{1}{c|}{91.61 $\pm$ 1.92} &
  \multicolumn{1}{c|}{86.14 $\pm$ 1.57} &
  \multicolumn{1}{c|}{8.0 $\pm$ 3.8} &
  24.2 $\pm$ 23.5 \\  \hline
TAO-Oblique &
  \multicolumn{1}{c|}{97.11 $\pm$ 0.59} &
  \multicolumn{1}{c|}{\textit{\textbf{95.47 $\pm$ 1.06}}} &
  \multicolumn{1}{c|}{\textit{\textbf{6.8 $\pm$ 1.2}}} &
 \textit{ \textbf{18.8 $\pm$ 5.8}} &
  \multicolumn{1}{c|}{89.90 $\pm$ 1.56} &
  \multicolumn{1}{c|}{\textit{\textbf{87.87 $\pm$ 1.50}}} &
  \multicolumn{1}{c|}{\textit{\textbf{5.2 $\pm$ 2.2}}} &
  \textit{\textbf{13.0 $\pm$ 7.1}} \\ \hline
\end{tabular}%
}
\end{table*}

\textit{\textbf{Topology setup}} is done with 15 multi-radio end nodes and one gateway. LoRa radios form a single-hop network where a LoRa radio can directly communicate with the gateway and Zigbee radios form a multi-hop mesh topology to reach the gateway.  Multi-hop Zigbee network uses a
distance-vector routing~\cite{cheng1989loop}
protocol for multi-hop routing. Since our region of interest is the \textit{gray-region} \cite{sundaram2024mars}, most of the multi-radio nodes are populated in the \textit{gray-region} (0.5-1.2Km from gateway) at two different locations A (Fig. \ref{fig:LocA-mesh-topo}) and B (Fig. \ref{fig:LocB-mesh-topo}).

\textit{\textbf{Data Collection}} is done on the mesh topology deployed in both the above-mentioned locations. Data is collected from the nodes populated in the \textit{gray-region} \cite{sundaram2024mars}. A total of 25,500 data packets were recorded. This comprehensive data set covers all the different dynamics of the deployed environment. The throughput of each transmitted data packet is recorded. This data set is manually labeled by a human to identify the high-throughput radio and its associated loss for each transmitted data packet. The associated loss is calculated as the difference in the throughput of a high and low-throughput radio. We formulate this radio selection problem as a classification problem and develop multiple classification models, that take in the input feature vector engineered with domain knowledge, following the same procedure as done in MARS \cite{sundaram2024mars}.

\textit{\textbf{Problem formulation, Feature Selection and Engineering.}} We formulated this radio selection problem as a classification problem and developed multiple classification models, that take in the input feature vector engineered with domain knowledge, following the same procedure as done in MARS \cite{sundaram2024mars}.

The classification model takes the input features 
E2E path quality of LoRa radios (E2E-$PQ_{LoRa}$), and the E2E path quality of Zigbee radios (E2E-$PQ_{Zigbee}$) to output a high throughput radio. The input feature vector can be expressed as:

\begin{equation}
  \label{eq:input_vector_mod}
  Input_{i} = [HN_{Zigbee},E2E\_RSSI_{LoRa}, E2E\_PRR_{Zigbee}, E2E\_RNP_{Zigbee}]        
\end{equation}

The output of the machine learning model is the radio predicted to have higher instantaneous throughput. This can be expressed as:

\begin{equation}
  \label{eq:output_vector}
  Output_{i} = [Zigbee | LoRa]  
\end{equation}

HN is the Hop Number that denotes a node's distance from the gateway in terms of hops. This value is obtained from the multi-hop Zigbee radios running a DV routing protocol \cite{cheng1989loop} for multi-hop Zigbee communications. $E2E\_RSSI_{LoRa}$ is the End-to-End RSSI of LoRa radios. This is obtained by exploiting channel reciprocity as done in MARS\cite{sundaram2024mars}. $E2E\_PRR_{Zigbee}$ is the end-to-end packet reception ratio of Zigbee radios and $E2E\_RNP_{Zigbee}$ is the Required Number of Packets \cite{cerpa2005temporal} obtained from Zigbee radios. 

\subsection{Cost-weighted accuracy metric}

The decision trees use the traditional 0/1 accuracy to calculate the train/test accuracy of the models. Since we develop models that are optimized to avert all the high-cost errors, we use the Cost-Weighted Accuracy (CWA) metric as explained below:

\begin{equation}
\label{eq:CWA}
    CWA=\frac{\sum_{n}c_nI(y_n, T(x_n))}{\sum_{n}c_n}.100
\end{equation}

where c is the cost. In the radio selector problem, the cost is the throughput loss incurred as the result of choosing a low-throughput radio instead of a high-throughput radio. I(p,q) is an indicator function that is equal to one when p = q, $y_n$ is the ground truth, T($x_n$) is the classification model that takes input $x_n$ and returns an output. 

\subsection{Models and their prediction accuracies}
\label{sec:ML_model}

The test and train CWA of three widely used models SVM \cite{cortes1995support}, Logistic Regression \cite{friedman2009elements} and CART \cite{praagman1985classification} decision trees are tabulated in Table \ref{tab:prediction_models}. These models are trained with 1500 samples like MARS~\cite{sundaram2024mars}.

\textbf{Logistic regression} (LR) \cite{friedman2009elements} is a widely used binary classification model that predicts the probability \( P(Y=1|X) \) using the logistic function applied to a linear predictor. The model takes the form:

\[
P(Y=1|X) = \frac{1}{1 + e^{-(\beta_0 + \beta_1 X_1 + \cdots + \beta_p X_p)}}
\]

where \( \beta_0, \beta_1, \dots, \beta_p \) are parameters estimated via maximum likelihood to optimize classification accuracy. By transforming the linear predictor to fit between 0 and 1, logistic regression is interpretable, can include regularization, and is adaptable to multi-class settings, making it suitable for a broad range of applications.

\textbf{Support Vector Machines} (SVM) \cite{cortes1995support} are supervised learning models used for classification that aim to find the optimal hyperplane separating two classes. Given training data \((x_i, y_i)\), where \( y_i \in \{-1, 1\} \), SVMs maximize the margin between the classes by solving:

\[
\min_{\mathbf{w}, b} \frac{1}{2} ||\mathbf{w}||^2
\]

subject to the constraints \( y_i(\mathbf{w} \cdot \mathbf{x}_i - b) \geq 1 \) for all \(i\). Here, \( \mathbf{w} \) represents the weight vector perpendicular to the hyperplane, and \( b \) is the bias. By maximizing the margin, SVMs achieve high generalization in high-dimensional spaces. In cases of non-linear separability, kernel functions can be employed to map data into a higher-dimensional space for linear separation.

\begin{table*}[t]
\centering
\caption{Decision rules of the tree shown in Fig. \ref{fig:tree_stability} for the decision nodes.}
\resizebox{\textwidth}{!}{%
\begin{tabular}{|c|c|ccccc|}
\hline
\multicolumn{1}{|l|}{\multirow{2}{*}{Tree Node IDs}} &
  \multicolumn{1}{l|}{\multirow{2}{*}{Training Data Size}} &
  \multicolumn{5}{c|}{Weights for linear combination of the features} \\ \cline{3-7} 
\multicolumn{1}{|l|}{} &
  \multicolumn{1}{l|}{} &
  \multicolumn{1}{c|}{\textbf{HN}} &
  \multicolumn{1}{c|}{\textbf{RSSI}} &
  \multicolumn{1}{c|}{\textbf{PRR}} &
  \multicolumn{1}{c|}{\textbf{RNP}} &
  \textbf{\begin{tabular}[c]{@{}c@{}}Constant\end{tabular}} \\
  \hline
\multirow{3}{*}{\textbf{Node 0}} &
  50\% &
  \multicolumn{1}{c|}{3.248390198} &
  \multicolumn{1}{c|}{0.155561545} &
  \multicolumn{1}{c|}{0} &
  \multicolumn{1}{c|}{-0.337416304} &
  -0.916876272 \\ \cline{2-7} 
 &
  75\% &
  \multicolumn{1}{c|}{3.248390198} &
  \multicolumn{1}{c|}{0.155561545} &
  \multicolumn{1}{c|}{0} &
  \multicolumn{1}{c|}{-0.337416304} &
  -0.916876272 \\ \cline{2-7} 
 &
  100\% &
  \multicolumn{1}{c|}{2.015888471} &
  \multicolumn{1}{c|}{-0.103138154} &
  \multicolumn{1}{c|}{-1.294708739} &
  \multicolumn{1}{c|}{-0.894449004} &
  -0.597743062 \\ \hline
\multirow{3}{*}{\textbf{Node 1}} &
  50\% &
  \multicolumn{1}{c|}{-4.028050107} &
  \multicolumn{1}{c|}{2.047812795} &
  \multicolumn{1}{c|}{-3.969356538} &
  \multicolumn{1}{c|}{0.147473355} &
  -6.011321658 \\ \cline{2-7} 
 &
  75\% &
  \multicolumn{1}{c|}{0.18954611} &
  \multicolumn{1}{c|}{1.925335047} &
  \multicolumn{1}{c|}{0} &
  \multicolumn{1}{c|}{2.371256232} &
  -1.477825092 \\ \cline{2-7} 
 &
  100\% &
  \multicolumn{1}{c|}{-0.829923334} &
  \multicolumn{1}{c|}{1.080856976} &
  \multicolumn{1}{c|}{-0.355732094} &
  \multicolumn{1}{c|}{1.376715261} &
  -0.45586578 \\ \hline
\multirow{3}{*}{\textbf{Node 2}} &
  50\% &
  \multicolumn{1}{c|}{0} &
  \multicolumn{1}{c|}{-0.848651273} &
  \multicolumn{1}{c|}{0} &
  \multicolumn{1}{c|}{-11.60563555} &
  -4.867864665 \\ \cline{2-7} 
 &
  75\% &
  \multicolumn{1}{c|}{-0.123607292} &
  \multicolumn{1}{c|}{-1.432806166} &
  \multicolumn{1}{c|}{-1.699720024} &
  \multicolumn{1}{c|}{-3.3523125} &
  -0.495526567 \\ \cline{2-7} 
 &
  100\% &
  \multicolumn{1}{c|}{-1.609863587} &
  \multicolumn{1}{c|}{-1.556952441} &
  \multicolumn{1}{c|}{1.217637854} &
  \multicolumn{1}{c|}{0.453619297} &
  -0.354708759 \\ \hline
\multirow{3}{*}{\textbf{Node 4}} &
  50\% &
  \multicolumn{1}{c|}{0} &
  \multicolumn{1}{c|}{-3.021387471} &
  \multicolumn{1}{c|}{0} &
  \multicolumn{1}{c|}{3.24466922} &
  8.311745459 \\ \cline{2-7} 
 &
  75\% &
  \multicolumn{1}{c|}{0} &
  \multicolumn{1}{c|}{-2.6919766} &
  \multicolumn{1}{c|}{0.928401407} &
  \multicolumn{1}{c|}{-2.224804842} &
  7.481391343 \\ \cline{2-7} 
 &
  100\% &
  \multicolumn{1}{c|}{-4.277725635} &
  \multicolumn{1}{c|}{-0.34558849} &
  \multicolumn{1}{c|}{2.282444039} &
  \multicolumn{1}{c|}{-1.167634787} &
  0.271872145 \\ \hline
\multirow{3}{*}{\textbf{Node 7}} &
  50\% &
  \multicolumn{1}{c|}{3.42766465} &
  \multicolumn{1}{c|}{6.744157135} &
  \multicolumn{1}{c|}{-6.236782959} &
  \multicolumn{1}{c|}{0.934803624} &
  -22.8890629 \\ \cline{2-7} 
 &
  75\% &
  \multicolumn{1}{c|}{1.992443923} &
  \multicolumn{1}{c|}{5.380224597} &
  \multicolumn{1}{c|}{-5.41441712} &
  \multicolumn{1}{c|}{1.699102024} &
  -22.83672636 \\ \cline{2-7} 
 &
  100\% &
  \multicolumn{1}{c|}{1.992443923} &
  \multicolumn{1}{c|}{5.380224597} &
  \multicolumn{1}{c|}{-5.41441712} &
  \multicolumn{1}{c|}{1.699102024} &
  -22.83672636 \\ \hline
\multirow{3}{*}{\textbf{Node 8}} &
  50\% &
  \multicolumn{1}{c|}{-4.549346595} &
  \multicolumn{1}{c|}{-5.351156989} &
  \multicolumn{1}{c|}{0.0875965} &
  \multicolumn{1}{c|}{-5.751555971} &
  5.460221236 \\ \cline{2-7} 
 &
  75\% &
  \multicolumn{1}{c|}{-2.203418965} &
  \multicolumn{1}{c|}{-3.659680033} &
  \multicolumn{1}{c|}{2.521893084} &
  \multicolumn{1}{c|}{-2.637760746} &
  6.127699317 \\ \cline{2-7} 
 &
  100\% &
  \multicolumn{1}{c|}{-4.874210512} &
  \multicolumn{1}{c|}{-2.144528513} &
  \multicolumn{1}{c|}{2.704740697} &
  \multicolumn{1}{c|}{-1.99839194} &
  1.930497302 \\ \hline
\end{tabular}%
}
\label{tab:decision_rules}
\end{table*}
The accuracies of these models are modified to account for loss as shown in equation~\ref{eq:CWA}. The training and testing accuracies averaged over a five-fold cross-validation are tabulated in Table \ref{tab:prediction_models}. SVM shows a higher testing accuracy than the training accuracy. This means the model is not well-generalized for unseen data. While Logistic Regression shows a similar trend in Location A, its training and testing accuracies are very low for Location B. The CART model achieves better results than SVM and LR. However, the difference between training and testing CWAs of CART is high, meaning that the model is not generalized for unseen data. Optimizing the Axis-aligned (CART) tree with TAO \cite{carreira2018alternating} solves this problem, but the Oblique tree \cite{hada2024sparse} optimized with TAO algorithm \cite{carreira2018alternating} achieves better CWA than axis-aligned trees. TAO-Oblique trees achieve lesser training accuracy than CART in Location B, but the testing CWA is higher than CART. This means that the TAO \cite{carreira2018alternating} algorithm optimizes the decision tree and makes it suitable for real-world deployments. On average, the TAO \cite{carreira2018alternating} algorithm can reduce the space complexity of the trees by 50\%. TAO \cite{carreira2018alternating} algorithm can optimize the decision trees trained with lesser training data to perform better in the real world while reducing the space complexity of the trees to be deployed. This makes the TAO \cite{carreira2018alternating} algorithm more suitable for IoT deployments requiring less data collection effort. So, we use TAO-oblique trees for our experiments. 

\section{TAO's structural stability property}
The traditional algorithms for learning decision trees, those based on greedy top-down induction, have the fundamental problem of producing very different models for different samples of the training set. This poses a significant limitation for learning a single interpretable decision tree. Practitioners trying to understand the fundamental essence of the problem might lose trust in the model if it keeps changing drastically for each small change in the training dataset. This problem applies to both axis-aligned and oblique decision trees\cite{hada2024sparse}. We overcome this issue by exploiting the stability property of the TAO \cite{carreira2018alternating} algorithm. The TAO algorithm can take an initial tree and optimize it for the newly changed dataset, i.e., it has a warm-starting ability in optimization \cite{li2002instability, turney1995bias}. 

\begin{figure}[t]
   \centering
   \includegraphics[scale=0.6]{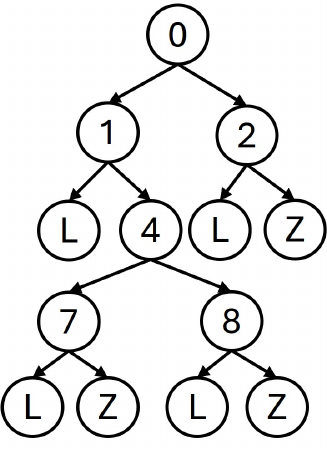}
        \caption{Tree structure provided by TAO for a tree model trained with 50\%, 75\% and 100\% of the training data.}
        \label{fig:tree_stability}
\end{figure}
\vspace{-0.05cm}
 \emph{TAO optimization provides a stable tree structure for any tree (Axis-aligned, Oblique, Sparse-Oblique \cite{hada2024sparse}) algorithm, even when new data is added to the oblique decision trees.} We experimented to quantify this stability property with our data set. During this experiment, we trained and optimized an oblique decision tree with 50\%, 75\%, and 100\% of the data samples from our training set. As expected, the TAO optimization algorithm provided a stable tree structure as shown in Figure \ref{fig:tree_stability}. 
\begin{table*}[t]
\centering
\caption{Decision rules of the sparse tree shown in Fig. \ref{fig:sparse_tree}.}
\resizebox{\textwidth}{!}{%
\begin{tabular}{|c|ccccc|}
\hline
\multicolumn{1}{|l|}{\multirow{2}{*}{Tree Node IDs}} &
  \multicolumn{5}{c|}{Weights for linear combination of the features} \\ \cline{2-6} 
\multicolumn{1}{|l|}{} &
  \multicolumn{1}{c|}{\textbf{HN}} &
  \multicolumn{1}{c|}{\textbf{RSSI}} &
  \multicolumn{1}{c|}{\textbf{PRR}} &
  \multicolumn{1}{c|}{\textbf{RNP}} &
  \textbf{\begin{tabular}[c]{@{}c@{}}Regularization \\ Constant\end{tabular}} \\ \hline
\textbf{Node 0} &
  \multicolumn{1}{c|}{0.958733267484} &
  \multicolumn{1}{c|}{1.025309937395} &
  \multicolumn{1}{c|}{-0.074761701442} &
  \multicolumn{1}{c|}{0.077471341337} &
  -0.122153674469 \\ \hline
\textbf{Node 1} &
  \multicolumn{1}{c|}{1.415681867042} &
  \multicolumn{1}{c|}{0} &
  \multicolumn{1}{c|}{0} &
  \multicolumn{1}{c|}{0} &
  0.143560158843 \\ \hline
\textbf{Node 4} &
  \multicolumn{1}{c|}{0} &
  \multicolumn{1}{c|}{-1.298779855192} &
  \multicolumn{1}{c|}{0} &
  \multicolumn{1}{c|}{-0.211013927858} &
  -1.448720963148 \\ \hline
\end{tabular}%
}
\label{tab:sparse_weights}
\end{table*}
There is a slight change in the weights of the decision rules when new data samples are added. In Figure \ref{fig:tree_stability}, numbered nodes are decision nodes, and lettered nodes are leaf nodes denoting the radio to be chosen for transmitting the packet (L for LoRa and Z for Zigbee). Since we are using oblique decision trees, the decision rules are a linear combination of the feature vectors with weights and a regularization constant as shown in Table 
\begin{figure}[t]
   \centering
   \includegraphics[scale=0.6]{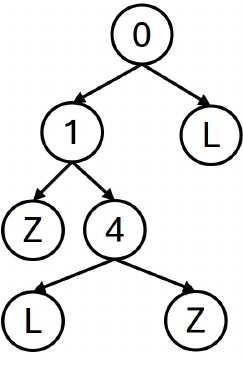}
        \caption{Sparse oblique tree}
        \label{fig:sparse_tree}
\end{figure}\ref{tab:decision_rules}. Figure \ref{fig:tree_stability} shows the tree structure obtained for training data sets of different sizes. For the training dataset of size 50\%, 75\%, and 100\% the test errors are 18.23\%, 17.85\%, and 14.41\% respectively. As expected, the test errors decrease as the training dataset size increases. However, the tree structure is always the same irrespective of retraining the model with new data samples because of the stability property of the TAO\cite{carreira2018alternating} algorithm.

Table \ref{tab:decision_rules} shows the decision rules of each decision node shown in Figure \ref{fig:tree_stability}. The decision rules are the linear combination of the features along with their weights and a regularization constant. When the node has to transmit a packet, it will input the input feature vector to the TAO-oblique decision tree. The TAO oblique decision tree will calculate this linear combination. If the result of this linear combination is negative, it traverses to the left child. If the result is positive, it traverses to the right child. This process is continued until a leaf node is reached. 

\section{Understanding radio selection via TAO's structural stability property}

We leverage TAO's\cite{carreira2018alternating} structural stability property to understand the essence of the radio selection problem. First, we show that we can interpret TAO-Oblique trees, but it is hard to extract a meaningful insight into the problem since a linear combination of all the features is used. So, we employ TAO-optimized sparse oblique trees to extract meaningful insights from the interpretations. 

The TAO-oblique tree is shown in Figure \ref{fig:tree_stability}. Each decision node (numbered node) in this tree uses a weighted linear combination of the features. The weights are shown in Table \ref{tab:decision_rules}. Node 0 is the root node where the entire dataset is divided into two. At Node 0, more emphasis is given to the feature \textit{Hop Number}. The hop number is the distance from a given node to the gateway. This is obtained using the distance vector protocol run by the Zigbee radios. Higher emphasis on the Hop number for Node 0 tells that the decision tree is choosing the node's distance from the gateway for the initial division of the dataset. 

Node 1 has one leaf node that decides to select the LoRa radio for transmission. If the linear combination for radios at Node 1 is negative, LoRa radio is chosen. At this Node 1, the decision tree gives more emphasis to RSSI and RNP. If the RSSI is higher, LoRa is chosen or the control goes to Node 4. Node 2 is a decision node with similar emphasis for every feature. 

Node 4 further divides into two more decision nodes 7 and, 8 which finally choose the radio for selection. This trend shows that using a threshold based on distance is not the only required factor for radio selection and further more details are required for accurately predicting the high-throughput radio. At this node, RNP and RSSI play an important role in dividing the dataset. At Nodes 7 and 8, all the features are given similar emphasis for radio selection.

From this structure, it is evident that the Hop Number (distance from the gateway) is used only for coarse-grained division of the dataset. The decision nodes that predict the high-throughput radio, use all the features (i.e.) fine-grained division of the dataset needs all the features to make a decision. It is hard to obtain a meaningful interpretation from this TAO-Oblique tree because the decision nodes use a linear combination of all the input features. We leverage Sparse-oblique trees to overcome this problem.    

TAO-optimized sparse-oblique trees enable the extraction of meaningful insights from the interpretations by reducing the number of features used at the decision nodes \cite{hada2024sparse}. Different combinations of all the feature vectors are tried non-homogeneously at each decision node. A tree providing better insights with reasonable accuracy is utilized for interpretations. The structure of the TAO-optimized sparse oblique tree is shown in Figure~\ref{fig:sparse_tree} and the corresponding weights are shown in Table~\ref{tab:sparse_weights}.

At Node 0, higher emphasis is given to the Hop number and RSSI to divide the dataset into two parts. If the data sample has a higher hop number, i.e. farther away from the gateway, with the best RSSI, the tree always chooses LoRa. 

At Node 1, we have the data samples that are not farther away from the gateway with better RSSI. This decision node emphasizes only on the Hop number.
\begin{figure*}[t]
  \captionsetup[subfigure]{justification=Centering}
  \centering
    \begin{subfigure}{0.49\linewidth}
     \centering
    \includegraphics[ height=7cm]{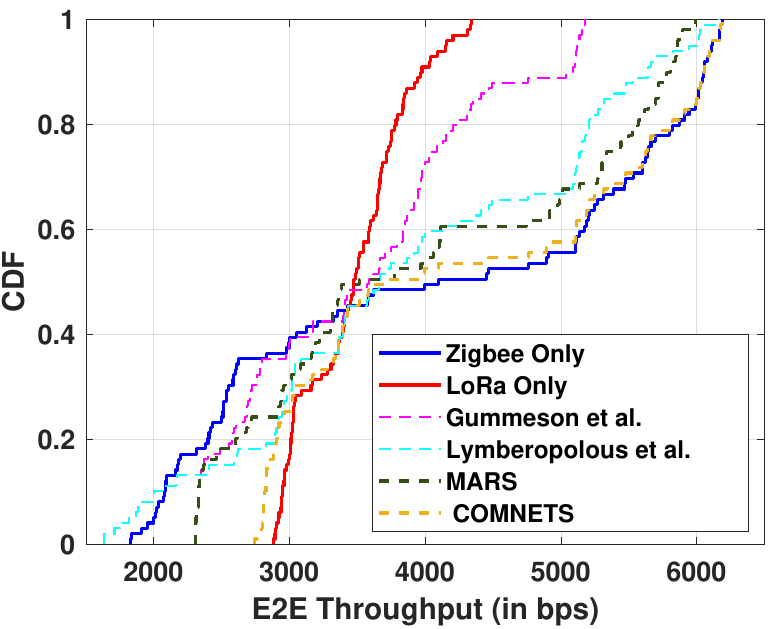}
    \caption{Throughput - Location A}
    \label{fig:locA-cdf}
  \end{subfigure}
  \begin{subfigure}{0.49\linewidth}
   \centering
    \includegraphics[ height=7cm]{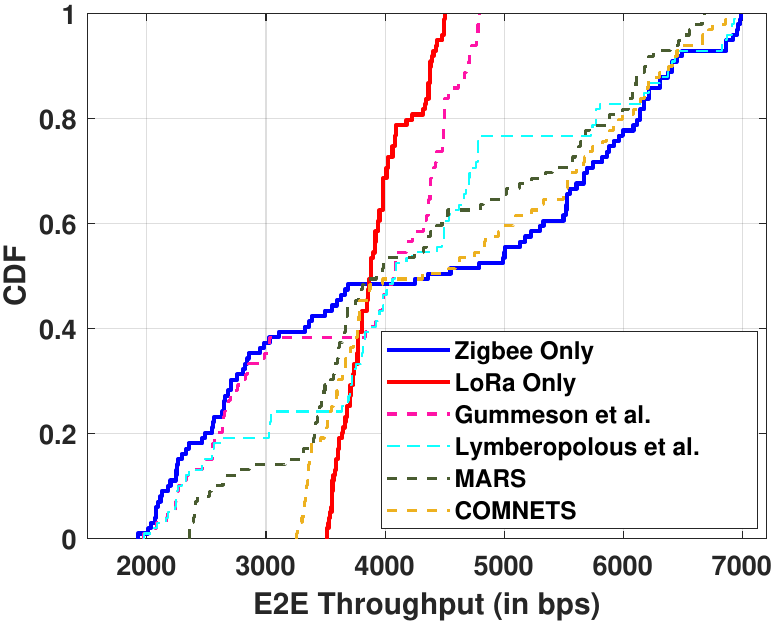}
    \caption{Throughput - Location B}
    \label{fig:locB-cdf}
  \end{subfigure}
  \caption{Throughput gain of \Name}
  \label{fig:throughput_gain}
\end{figure*}
\begin{figure*}[t]
    \centering
    \includegraphics[width=0.8\linewidth, height = 5.5cm]{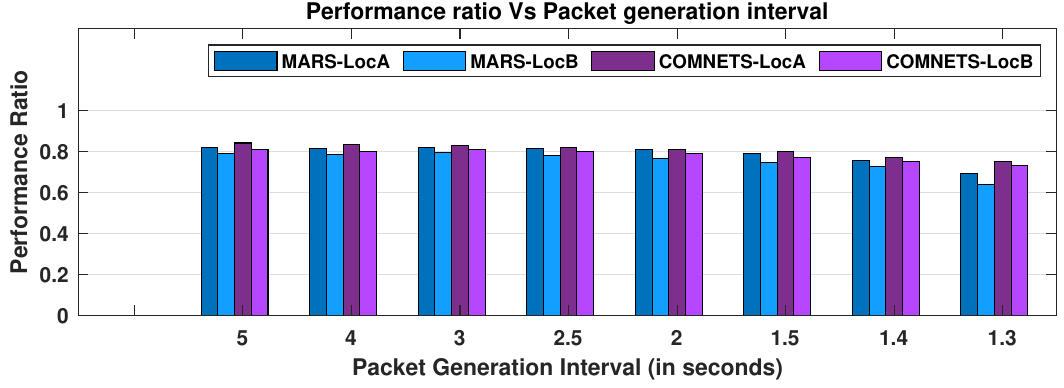}
    \caption{Performance Ratio on different packet generation intervals}
    \label{fig:pack-break}
\end{figure*}
If the hop number is lower, i.e., the nodes are closer to the gateway. The tree always chooses the Zigbee radio to transmit as it provides higher throughput at closer distances.  

At Node 4, we have the data samples that are in between farther and closer nodes. At this distance, higher emphasis is given to the RSSI of the LoRa radio and RNP \cite{cerpa2005temporal} of the Zigbee radio. This means that RNP and RSSI are the two important features required to identify the high-throughput radio in the gray-region.  Note that PDR is not as useful as RNP for the multi-hop Zigbee radios.  We infer that this may be the case due to the RNP metric providing a quantitative measurement of the underlying distribution of packet losses instead of just an average number like PDR, leading to a better estimate of expected throughput.

\section{Large-scale evaluations}
\label{sec:experiments}
\Name\xspace is evaluated on a real-world, large-scale mesh topology at two different locations as shown in Figures \ref{fig:LocA-mesh-topo} and \ref{fig:LocB-mesh-topo}. The deployments were done in complex environments that consisted of different building materials, human influx, and fleeting reflectors. The end devices will schedule a packet every three seconds periodically, destined to the gateway. When the packet is scheduled, the TAO\cite{carreira2018alternating}-oblique tree-based radio selector will select a radio for transmission. 10,400 data packets were transmitted for these experiments.   

\textbf{\textit{Benchmarks.}} The MARS \cite{sundaram2024mars} multi-radio system already outperforms Gummeson et al. \cite{gummeson2009adaptive} and the threshold-based~\cite{lymberopoulos2008towards} multi-radio system. However, we compare \Name\xspace with LoRa and Zigbee single radio systems and all the above-mentioned multi-radio systems. 

\textbf{\textit{Metrics.}} We evaluated the performance of \Name\xspace at two different locations on the following metrics: (i) Throughput, (ii) Latency, and performance ratio on different packet generation intervals. 

\begin{figure*}[t]
    \centering
    \includegraphics[width=0.8\linewidth, height = 6cm]{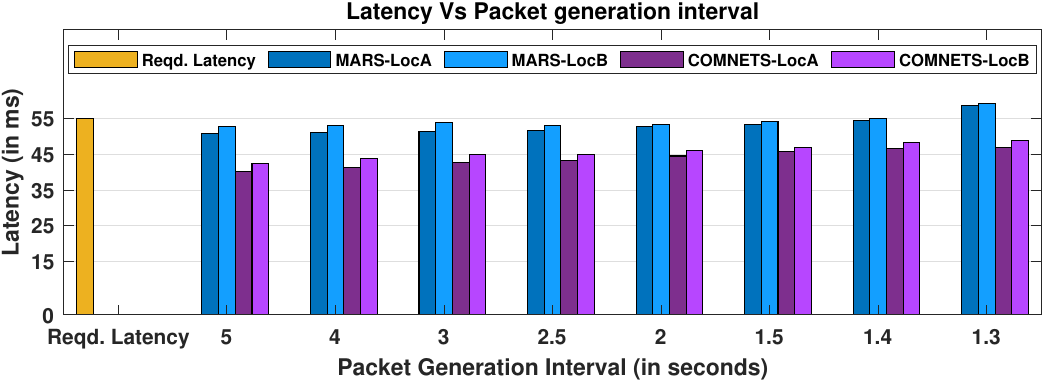}
    \caption{Latency on different packet generation intervals}
    \label{fig:latency}
\end{figure*}
\textbf{\textit{Results.}} \emph{The throughput gain} of \xspace\Name\xspace is plotted in Figure \ref{fig:locA-cdf} and \ref{fig:locB-cdf} for both the locations A and B. At Location A (Fig. \ref{fig:locA-cdf}), Zigbee achieves high throughput for 42\% of the transmissions and LoRa achieves higher throughput for the rest of the transmissions. MARS follows the high throughput radio all the time with slight errors for the first 20\% of the transmissions and for the transmission between 50-55\% of the transmissions. This is happening because MARS considers both high-throughput and low-throughput errors to be the same. At both locations, MARS outperforms Gummeson et al. \cite{gummeson2009adaptive} and Lymberopolous et al. \cite{lymberopoulos2008towards}. At Location B,  Gummeson et al. \cite{gummeson2009adaptive} and Lymberopolous et al. \cite{lymberopoulos2008towards} tend to cross the solid red line. Theoretically, this is impossible as the multi-radio systems cannot cross the performance of the single radio system. However, In practice, these systems were evaluated at different times, when the channel quality (uncontrollable factor) would be different. This leads to a small jitter in throughput, making the performance go beyond the single-radio system. \Name\xspace optimizing to avoid high-throughput errors achieves an average throughput gain of 20.83\% than MARS. Similar trends are seen in Location B (Fig. \ref{fig:locB-cdf}). Here, \Name\xspace achieves an average throughput gain of 17.39\% over MARS.  

\textbf{\emph{Performance Ratio of \xspace\Name\xspace on different packet generation intervals}} is plotted in Figure \ref{fig:pack-break}. The performance ratio is calculated as the average per-packet throughput of \Name\xspace over the average per-packet throughput of the best radio. In this figure, we can see that the performance ratio of MARS slightly decreases with a decrease in packet generation interval from 5-1.5s. A significant decrease in the performance of MARS happens for the packet generation intervals 1.4s and 1.3s. This significant decrease is due to the queuing delays caused by the increase in data packets. This queuing delay also affects the beacon packets that are used for path quality estimation and the packets that carry this information to other nodes to calculate path quality. We follow the same path quality estimation as done in MARS \cite{sundaram2024mars}. The high-cost errors happening due to this phenomenon cause this significant dip in performance ratio. \Name\xspace follows similar trends with a decrease in the packet generation interval. However, \Name\xspace powered by TAO\cite{carreira2018alternating}-oblique trees, optimized to avert high-cost errors, achieves a higher performance ratio than MARS as it can avert most of the high-throughput errors. This cost-sensitive optimization helps to improve the overall performance of multi-radio networks even with inevitable detriments.

\textbf{\emph{Latency of \xspace\Name\xspace on different packet generation intervals}} is plotted in Figure \ref{fig:latency}. According to 5G America's report~\cite{5Gamericas}, the average required latency for mesoscale IoT applications is 55ms. The latency of MARS slightly increases with a decrease in packet generation interval. The latency of MARS goes beyond the required 55ms when the packet generation interval is 1.3s. This is because MARS does not optimize for high-cost errors. However, \xspace\Name\xspace powered by TAO\cite{carreira2018alternating}-Oblique trees optimized to avert high-cost errors can keep the latency of the system within the required bounds.  

\vspace{-0.1cm}
\section{Conclusion and future work}
We presented \Name, a cost-sensitive ML-based radio selection algorithm to predict high throughput radio for mesoscale IoT applications. We showed that MARS suffers from the problem of cost sensitivity, giving a greater margin for throughput optimization. The current cost-sensitive ML algorithms set a misclassification cost for each class while the radio selection problem for multi-radio networks entails different misclassification costs for each data sample. Optimizing such cost-sensitive costs is tedious. First, we leverage the TAO algorithm to overcome this issue. TAO not only optimizes the cost for each data sample, it also optimizes the tree for unseen data along with size reduction, making it efficient for deployment in resource-constrained IoT devices. Second, we leverage the structural stability property of TAO to understand the essence of the radio selection problem. By using TAO-optimized sparse oblique trees, we gain a more fundamental understanding of the importance of the different input features, and how \Name\xspace makes the classification decisions at different regions in the network.  Having a tree structure allows for easier interpretability than other ML techniques (e.g. neural networks). Finally, our evaluations on large-scale, real-world deployments show that \Name\xspace achieves a throughput gain of 20.83\% and 17.39\% than MARS at two locations A and B respectively.  

In the future, we would also like to explore different input features for the decision trees.  In particular, we would like to try to feed raw data packet sequences instead of path-quality estimation metrics, and let the model find patterns not easily detected by a human designer on the raw data. On the hardware aspect, the multi-radio hardware used still has a large room for improvement. For example, we would like to explore different hardware requirements for running different ML algorithms and optimize them to enable quicker on-board inference and re-training of the model. This will facilitate the development of AI/ML-based protocols for IoT networking. In the modeling aspect, we would like to deploy the system for a longer period to evaluate how the ML model works for different seasons over the year. This will give an insight into the required retraining interval in the beginning and how the retraining requirement flattens out over time. 

\balance
\bibliographystyle{ACM-Reference-Format}
\bibliography{sample-base}


\begin{thebibliography}{40}


\ifx \showCODEN    \undefined \def \showCODEN     #1{\unskip}     \fi
\ifx \showDOI      \undefined \def \showDOI       #1{#1}\fi
\ifx \showISBNx    \undefined \def \showISBNx     #1{\unskip}     \fi
\ifx \showISBNxiii \undefined \def \showISBNxiii  #1{\unskip}     \fi
\ifx \showISSN     \undefined \def \showISSN      #1{\unskip}     \fi
\ifx \showLCCN     \undefined \def \showLCCN      #1{\unskip}     \fi
\ifx \shownote     \undefined \def \shownote      #1{#1}          \fi
\ifx \showarticletitle \undefined \def \showarticletitle #1{#1}   \fi
\ifx \showURL      \undefined \def \showURL       {\relax}        \fi
\providecommand\bibfield[2]{#2}
\providecommand\bibinfo[2]{#2}
\providecommand\natexlab[1]{#1}
\providecommand\showeprint[2][]{arXiv:#2}

\bibitem[Al~Islam et~al\mbox{.}(2011)]%
        {al2011backpacking}
\bibfield{author}{\bibinfo{person}{ABM~Alim Al~Islam}, \bibinfo{person}{Mohammad~S Hossain}, \bibinfo{person}{Vijay Raghunathan}, {and} \bibinfo{person}{Y~Charlie Hu}.} \bibinfo{year}{2011}\natexlab{}.
\newblock \showarticletitle{Backpacking: Deployment of heterogeneous radios in high data rate sensor networks}. In \bibinfo{booktitle}{\emph{2011 Proceedings of 20th international conference on computer communications and networks (ICCCN)}}. IEEE, \bibinfo{pages}{1--8}.
\newblock


\bibitem[Americas(2019)]%
        {5Gamericas}
\bibfield{author}{\bibinfo{person}{5G Americas}.} \bibinfo{year}{2019}\natexlab{}.
\newblock \bibinfo{booktitle}{\emph{{5G - The future of IoT}}}.
\newblock
\urldef\tempurl%
\url{http://tinyurl.com/26y5epsp}
\showURL{%
\tempurl}


\bibitem[Ananthanarayanan and Stoica(2009)]%
        {ananthanarayanan2009blue}
\bibfield{author}{\bibinfo{person}{Ganesh Ananthanarayanan} {and} \bibinfo{person}{Ion Stoica}.} \bibinfo{year}{2009}\natexlab{}.
\newblock \showarticletitle{Blue-Fi: enhancing Wi-Fi performance using bluetooth signals}. In \bibinfo{booktitle}{\emph{Proceedings of the 7th international conference on Mobile systems, applications, and services}}. \bibinfo{pages}{249--262}.
\newblock


\bibitem[Bahl et~al\mbox{.}(2004)]%
        {bahl2004reconsidering}
\bibfield{author}{\bibinfo{person}{Paramvir Bahl}, \bibinfo{person}{Atul Adya}, \bibinfo{person}{Jitendra Padhye}, {and} \bibinfo{person}{Alec Wolman}.} \bibinfo{year}{2004}\natexlab{}.
\newblock \showarticletitle{Reconsidering wireless systems with multiple radios}.
\newblock \bibinfo{journal}{\emph{ACM SIGCOMM Computer Communication Review}} \bibinfo{volume}{34}, \bibinfo{number}{5} (\bibinfo{year}{2004}), \bibinfo{pages}{39--46}.
\newblock


\bibitem[Cerpa et~al\mbox{.}(2005)]%
        {cerpa2005temporal}
\bibfield{author}{\bibinfo{person}{Alberto Cerpa}, \bibinfo{person}{Jennifer~L Wong}, \bibinfo{person}{Miodrag Potkonjak}, {and} \bibinfo{person}{Deborah Estrin}.} \bibinfo{year}{2005}\natexlab{}.
\newblock \showarticletitle{Temporal properties of low power wireless links: modeling and implications on multi-hop routing}. In \bibinfo{booktitle}{\emph{Proceedings of the 6th ACM international symposium on Mobile ad hoc networking and computing}}. \bibinfo{pages}{414--425}.
\newblock


\bibitem[Cheng et~al\mbox{.}(1989)]%
        {cheng1989loop}
\bibfield{author}{\bibinfo{person}{Chunhsiang Cheng}, \bibinfo{person}{Ralph Riley}, \bibinfo{person}{Srikanta~PR Kumar}, {and} \bibinfo{person}{Jose~J Garcia-Luna-Aceves}.} \bibinfo{year}{1989}\natexlab{}.
\newblock \showarticletitle{A loop-free extended Bellman-Ford routing protocol without bouncing effect}.
\newblock \bibinfo{journal}{\emph{ACM SIGCOMM Computer Communication Review}} \bibinfo{volume}{19}, \bibinfo{number}{4} (\bibinfo{year}{1989}), \bibinfo{pages}{224--236}.
\newblock


\bibitem[Cortes(1995)]%
        {cortes1995support}
\bibfield{author}{\bibinfo{person}{Corinna Cortes}.} \bibinfo{year}{1995}\natexlab{}.
\newblock \showarticletitle{Support-Vector Networks}.
\newblock \bibinfo{journal}{\emph{Machine Learning}} (\bibinfo{year}{1995}).
\newblock


\bibitem[Domingos(1999)]%
        {domingos1999metacost}
\bibfield{author}{\bibinfo{person}{Pedro Domingos}.} \bibinfo{year}{1999}\natexlab{}.
\newblock \showarticletitle{Metacost: A general method for making classifiers cost-sensitive}. In \bibinfo{booktitle}{\emph{Proceedings of the fifth ACM SIGKDD international conference on Knowledge discovery and data mining}}. \bibinfo{pages}{155--164}.
\newblock


\bibitem[Draves et~al\mbox{.}(2004)]%
        {draves2004routing}
\bibfield{author}{\bibinfo{person}{Richard Draves}, \bibinfo{person}{Jitendra Padhye}, {and} \bibinfo{person}{Brian Zill}.} \bibinfo{year}{2004}\natexlab{}.
\newblock \showarticletitle{Routing in multi-radio, multi-hop wireless mesh networks}. In \bibinfo{booktitle}{\emph{Proceedings of the 10th annual international conference on Mobile computing and networking}}. \bibinfo{pages}{114--128}.
\newblock


\bibitem[Electric(2024a)]%
        {mitsubishielectricLargecapacityBattery}
\bibfield{author}{\bibinfo{person}{Mitsubishi Electric}.} \bibinfo{year}{2024}\natexlab{a}.
\newblock \bibinfo{title}{{L}arge-capacity battery control system}.
\newblock \bibinfo{howpublished}{\url{http://tinyurl.com/2ryvbcxv}}.
\newblock


\bibitem[Electric(2024b)]%
        {mitsubishielectricSmartMeter}
\bibfield{author}{\bibinfo{person}{Mitsubishi Electric}.} \bibinfo{year}{2024}\natexlab{b}.
\newblock \bibinfo{title}{{S}mart meter system}.
\newblock \bibinfo{howpublished}{\url{http://tinyurl.com/4z2dy6nc}}.
\newblock


\bibitem[Electric(2024c)]%
        {mitsubishielectricPowerSystems}
\bibfield{author}{\bibinfo{person}{Mitsubishi Electric}.} \bibinfo{year}{2024}\natexlab{c}.
\newblock \bibinfo{title}{“{B}{L}{E}n{D}er®”}.
\newblock \bibinfo{howpublished}{\url{http://tinyurl.com/3k435ymu}}.
\newblock


\bibitem[Elkan(2001)]%
        {elkan2001foundations}
\bibfield{author}{\bibinfo{person}{Charles Elkan}.} \bibinfo{year}{2001}\natexlab{}.
\newblock \showarticletitle{The foundations of cost-sensitive learning}. In \bibinfo{booktitle}{\emph{International joint conference on artificial intelligence}}, Vol.~\bibinfo{volume}{17}. Lawrence Erlbaum Associates Ltd, \bibinfo{pages}{973--978}.
\newblock


\bibitem[et~al.(2018)]%
        {carreira2018alternating}
\bibfield{author}{\bibinfo{person}{Carreira-Perpin{\'a}n et al.}} \bibinfo{year}{2018}\natexlab{}.
\newblock \showarticletitle{Alternating optimization of decision trees, with application to learning sparse oblique trees}.
\newblock \bibinfo{journal}{\emph{In NeurIPS '18}} (\bibinfo{year}{2018}).
\newblock


\bibitem[Fan et~al\mbox{.}(2008)]%
        {liblinear}
\bibfield{author}{\bibinfo{person}{Rong-En Fan}, \bibinfo{person}{Kai-Wei Chang}, \bibinfo{person}{Cho-Jui Hsieh}, \bibinfo{person}{Xiang-Rui Wang}, {and} \bibinfo{person}{Chih-Jen Lin}.} \bibinfo{year}{2008}\natexlab{}.
\newblock \showarticletitle{LIBLINEAR: A Library for Large Linear Classification}.
\newblock \bibinfo{journal}{\emph{J. Mach. Learn. Res.}}  \bibinfo{volume}{9} (\bibinfo{date}{June} \bibinfo{year}{2008}), \bibinfo{pages}{1871–1874}.
\newblock
\showISSN{1532-4435}


\bibitem[Friedman(2009)]%
        {friedman2009elements}
\bibfield{author}{\bibinfo{person}{Jerome Friedman}.} \bibinfo{year}{2009}\natexlab{}.
\newblock \showarticletitle{The elements of statistical learning: Data mining, inference, and prediction}.
\newblock \bibinfo{journal}{\emph{(No Title)}} (\bibinfo{year}{2009}).
\newblock


\bibitem[Fu et~al\mbox{.}(2024)]%
        {fu2024comprehensive}
\bibfield{author}{\bibinfo{person}{Ruijie Fu}, \bibinfo{person}{Lancong Guo}, \bibinfo{person}{An Zou}, \bibinfo{person}{Cailian Chen}, \bibinfo{person}{Xinping Guan}, {and} \bibinfo{person}{Yehan Ma}.} \bibinfo{year}{2024}\natexlab{}.
\newblock \showarticletitle{Comprehensive Optimal Network Scheduling Strategies for Wireless Control Systems}.
\newblock \bibinfo{journal}{\emph{ACM Transactions on Cyber-Physical Systems}} (\bibinfo{year}{2024}).
\newblock


\bibitem[Glady et~al\mbox{.}(2009)]%
        {glady2009modeling}
\bibfield{author}{\bibinfo{person}{Nicolas Glady}, \bibinfo{person}{Bart Baesens}, {and} \bibinfo{person}{Christophe Croux}.} \bibinfo{year}{2009}\natexlab{}.
\newblock \showarticletitle{Modeling churn using customer lifetime value}.
\newblock \bibinfo{journal}{\emph{European journal of operational research}} \bibinfo{volume}{197}, \bibinfo{number}{1} (\bibinfo{year}{2009}), \bibinfo{pages}{402--411}.
\newblock


\bibitem[Gu et~al\mbox{.}(2019)]%
        {gu2019one}
\bibfield{author}{\bibinfo{person}{Chaojie Gu}, \bibinfo{person}{Rui Tan}, {and} \bibinfo{person}{Xin Lou}.} \bibinfo{year}{2019}\natexlab{}.
\newblock \showarticletitle{One-hop out-of-band control planes for multi-hop wireless sensor networks}.
\newblock \bibinfo{journal}{\emph{ACM Transactions on Sensor Networks (TOSN)}} \bibinfo{volume}{15}, \bibinfo{number}{4} (\bibinfo{year}{2019}), \bibinfo{pages}{1--29}.
\newblock


\bibitem[Gummeson et~al\mbox{.}(2009)]%
        {gummeson2009adaptive}
\bibfield{author}{\bibinfo{person}{Jeremy Gummeson}, \bibinfo{person}{Deepak Ganesan}, \bibinfo{person}{Mark~D Corner}, {and} \bibinfo{person}{Prashant Shenoy}.} \bibinfo{year}{2009}\natexlab{}.
\newblock \showarticletitle{An adaptive link layer for range diversity in multi-radio mobile sensor networks}. In \bibinfo{booktitle}{\emph{IEEE INFOCOM 2009}}. IEEE, \bibinfo{pages}{154--162}.
\newblock


\bibitem[Hada et~al\mbox{.}(2024)]%
        {hada2024sparse}
\bibfield{author}{\bibinfo{person}{Suryabhan~Singh Hada}, \bibinfo{person}{Miguel~{\'A} Carreira-Perpi{\~n}{\'a}n}, {and} \bibinfo{person}{Arman Zharmagambetov}.} \bibinfo{year}{2024}\natexlab{}.
\newblock \showarticletitle{Sparse oblique decision trees: A tool to understand and manipulate neural net features}.
\newblock \bibinfo{journal}{\emph{Data Mining and Knowledge Discovery}} \bibinfo{volume}{38}, \bibinfo{number}{5} (\bibinfo{year}{2024}), \bibinfo{pages}{2863--2902}.
\newblock


\bibitem[Jan et~al\mbox{.}(2012)]%
        {jan2012simple}
\bibfield{author}{\bibinfo{person}{Te-Kang Jan}, \bibinfo{person}{Da-Wei Wang}, \bibinfo{person}{Chi-Hung Lin}, {and} \bibinfo{person}{Hsuan-Tien Lin}.} \bibinfo{year}{2012}\natexlab{}.
\newblock \showarticletitle{A simple methodology for soft cost-sensitive classification}. In \bibinfo{booktitle}{\emph{Proceedings of the 18th ACM SIGKDD international conference on Knowledge discovery and data mining}}. \bibinfo{pages}{141--149}.
\newblock


\bibitem[Jin et~al\mbox{.}({[n.\,d.]})]%
        {jin2011wizi}
\bibfield{author}{\bibinfo{person}{Tao Jin}, \bibinfo{person}{Guevara Noubir}, {and} \bibinfo{person}{Bo Sheng}.} \bibinfo{year}{[n.\,d.]}\natexlab{}.
\newblock \showarticletitle{Wizi-cloud: Application-transparent dual zigbee-wifi radios for low power internet access}. In \bibinfo{booktitle}{\emph{IEEE INFOCOM, 2011}}.
\newblock


\bibitem[Kim et~al\mbox{.}(2012)]%
        {kim2012classification}
\bibfield{author}{\bibinfo{person}{Jungeun Kim}, \bibinfo{person}{Keunho Choi}, \bibinfo{person}{Gunwoo Kim}, {and} \bibinfo{person}{Yongmoo Suh}.} \bibinfo{year}{2012}\natexlab{}.
\newblock \showarticletitle{Classification cost: An empirical comparison among traditional classifier, Cost-Sensitive Classifier, and MetaCost}.
\newblock \bibinfo{journal}{\emph{Expert Systems with Applications}} \bibinfo{volume}{39}, \bibinfo{number}{4} (\bibinfo{year}{2012}), \bibinfo{pages}{4013--4019}.
\newblock


\bibitem[Kusy et~al\mbox{.}(2014)]%
        {kusy2014radio}
\bibfield{author}{\bibinfo{person}{Branislav Kusy}, \bibinfo{person}{David Abbott}, \bibinfo{person}{Christian Richter}, \bibinfo{person}{Cong Huynh}, \bibinfo{person}{Mikhail Afanasyev}, \bibinfo{person}{Wen Hu}, \bibinfo{person}{Michael Br{\"u}nig}, \bibinfo{person}{Diethelm Ostry}, {and} \bibinfo{person}{Raja Jurdak}.} \bibinfo{year}{2014}\natexlab{}.
\newblock \showarticletitle{Radio diversity for reliable communication in sensor networks}.
\newblock \bibinfo{journal}{\emph{ACM Transactions on Sensor Networks (TOSN)}} \bibinfo{volume}{10}, \bibinfo{number}{2} (\bibinfo{year}{2014}), \bibinfo{pages}{1--29}.
\newblock


\bibitem[Li and Belford(2002)]%
        {li2002instability}
\bibfield{author}{\bibinfo{person}{Ruey-Hsia Li} {and} \bibinfo{person}{Geneva~G Belford}.} \bibinfo{year}{2002}\natexlab{}.
\newblock \showarticletitle{Instability of decision tree classification algorithms}. In \bibinfo{booktitle}{\emph{Proceedings of the eighth ACM SIGKDD international conference on Knowledge discovery and data mining}}. \bibinfo{pages}{570--575}.
\newblock


\bibitem[Lin(2019)]%
        {lin2019advances}
\bibfield{author}{\bibinfo{person}{Hsuan-Tien Lin}.} \bibinfo{year}{2019}\natexlab{}.
\newblock \showarticletitle{Advances in cost-sensitive multiclass and multilabel classification}. In \bibinfo{booktitle}{\emph{Proceedings of the 25th ACM SIGKDD International Conference on Knowledge Discovery \& Data Mining}}. \bibinfo{pages}{3187--3188}.
\newblock


\bibitem[LLC({[n.\,d.]})]%
        {ronoth_llc}
\bibfield{author}{\bibinfo{person}{Ronoth LLC}.} \bibinfo{year}{[n.\,d.]}\natexlab{}.
\newblock \bibinfo{title}{{LoStik}}.
\newblock \bibinfo{howpublished}{\url{https://github.com/ronoth/lostik}}.
\newblock


\bibitem[Lymberopoulos et~al\mbox{.}(2008)]%
        {lymberopoulos2008towards}
\bibfield{author}{\bibinfo{person}{Dimitrios Lymberopoulos}, \bibinfo{person}{Nissanka~B Priyantha}, \bibinfo{person}{Michel Goraczko}, {and} \bibinfo{person}{Feng Zhao}.} \bibinfo{year}{2008}\natexlab{}.
\newblock \showarticletitle{Towards energy efficient design of multi-radio platforms for wireless sensor networks}. In \bibinfo{booktitle}{\emph{2008 International Conference on Information Processing in Sensor Networks (ipsn 2008)}}. IEEE, \bibinfo{pages}{257--268}.
\newblock


\bibitem[Nagai et~al\mbox{.}(2024)]%
        {nagai2024improve}
\bibfield{author}{\bibinfo{person}{Yukimasa Nagai}, \bibinfo{person}{Jianlin Guo}, \bibinfo{person}{Takenori Sumi}, \bibinfo{person}{Kieran Parsons}, \bibinfo{person}{Philip Orlik}, \bibinfo{person}{Benjamin~A Rolfe}, {and} \bibinfo{person}{Pu Wang}.} \bibinfo{year}{2024}\natexlab{}.
\newblock \showarticletitle{Improve IEEE 802.15. 4 Network Reliability by Suspendable CSMA/CA}. In \bibinfo{booktitle}{\emph{2024 IEEE Wireless Communications and Networking Conference (WCNC)}}. IEEE, \bibinfo{pages}{1--6}.
\newblock


\bibitem[Praagman(1985)]%
        {praagman1985classification}
\bibfield{author}{\bibinfo{person}{J Praagman}.} \bibinfo{year}{1985}\natexlab{}.
\newblock \bibinfo{title}{Classification and regression trees: Leo BREIMAN, Jerome H. FRIEDMAN, Richard A. OLSHEN and Charles J. STONE The Wadsworth Statistics/Probability Series, Wadsworth, Belmont, 1984, x+ 358 pages}.
\newblock
\newblock


\bibitem[Sundaram et~al\mbox{.}(2019)]%
        {sundaram2019survey}
\bibfield{author}{\bibinfo{person}{Jothi Prasanna~Shanmuga Sundaram}, \bibinfo{person}{Wan Du}, {and} \bibinfo{person}{Zhiwei Zhao}.} \bibinfo{year}{2019}\natexlab{}.
\newblock \showarticletitle{A survey on LoRa networking: Research problems, current solutions, and open issues}.
\newblock \bibinfo{journal}{\emph{IEEE Communications Surveys \& Tutorials}} \bibinfo{volume}{22}, \bibinfo{number}{1} (\bibinfo{year}{2019}), \bibinfo{pages}{371--388}.
\newblock


\bibitem[Sundaram et~al\mbox{.}(2024)]%
        {sundaram2024mars}
\bibfield{author}{\bibinfo{person}{Jothi Prasanna~Shanmuga Sundaram}, \bibinfo{person}{Arman Zharmagambetov}, \bibinfo{person}{Magzhan Gabidolla}, \bibinfo{person}{Miguel~A Carreira-Perpinan}, {and} \bibinfo{person}{Alberto Cerpa}.} \bibinfo{year}{2024}\natexlab{}.
\newblock \showarticletitle{MARS: Multi-radio Architecture with Radio Selection using Decision Trees for emerging mesoscale CPS/IoT applications}.
\newblock \bibinfo{journal}{\emph{arXiv preprint arXiv:2409.18043}} (\bibinfo{year}{2024}).
\newblock


\bibitem[Sur et~al\mbox{.}({[n.\,d.]})]%
        {sur2017wifi}
\bibfield{author}{\bibinfo{person}{Sanjib Sur}, \bibinfo{person}{Ioannis Pefkianakis}, \bibinfo{person}{Xinyu Zhang}, {and} \bibinfo{person}{Kyu-Han Kim}.} \bibinfo{year}{[n.\,d.]}\natexlab{}.
\newblock \showarticletitle{WiFi-assisted 60 GHz wireless networks}. In \bibinfo{booktitle}{\emph{ACM MobiCom, 2017}}.
\newblock


\bibitem[Tan et~al\mbox{.}(2018)]%
        {tan2018resilience}
\bibfield{author}{\bibinfo{person}{Rui Tan}, \bibinfo{person}{Linshan Jiang}, \bibinfo{person}{Arvind Easwaran}, {et~al\mbox{.}}} \bibinfo{year}{2018}\natexlab{}.
\newblock \showarticletitle{Resilience bounds of sensing-based network clock synchronization}. In \bibinfo{booktitle}{\emph{2018 IEEE 24th International Conference on Parallel and Distributed Systems (ICPADS)}}. IEEE, \bibinfo{pages}{894--902}.
\newblock


\bibitem[Technology(2021)]%
        {telosb}
\bibfield{author}{\bibinfo{person}{Crossbow Technology}.} \bibinfo{year}{2021}\natexlab{}.
\newblock \bibinfo{title}{{TelosB}}.
\newblock \bibinfo{howpublished}{\url{http://tinyurl.com/2svnbhjs}}.
\newblock


\bibitem[Turney(1995)]%
        {turney1995bias}
\bibfield{author}{\bibinfo{person}{Peter Turney}.} \bibinfo{year}{1995}\natexlab{}.
\newblock \showarticletitle{Bias and the quantification of stability}.
\newblock \bibinfo{journal}{\emph{Machine Learning}}  \bibinfo{volume}{20} (\bibinfo{year}{1995}), \bibinfo{pages}{23--33}.
\newblock


\bibitem[Vasisht et~al\mbox{.}(2017)]%
        {vasisht2017farmbeats}
\bibfield{author}{\bibinfo{person}{Deepak Vasisht}, \bibinfo{person}{Zerina Kapetanovic}, \bibinfo{person}{Jongho Won}, \bibinfo{person}{Xinxin Jin}, \bibinfo{person}{Ranveer Chandra}, \bibinfo{person}{Sudipta Sinha}, \bibinfo{person}{Ashish Kapoor}, \bibinfo{person}{Madhusudhan Sudarshan}, {and} \bibinfo{person}{Sean Stratman}.} \bibinfo{year}{2017}\natexlab{}.
\newblock \showarticletitle{$\{$FarmBeats$\}$: an $\{$IoT$\}$ platform for $\{$Data-Driven$\}$ agriculture}. In \bibinfo{booktitle}{\emph{14th USENIX Symposium on Networked Systems Design and Implementation (NSDI 17)}}. \bibinfo{pages}{515--529}.
\newblock


\bibitem[Yang et~al\mbox{.}(2024)]%
        {yang2024rateless}
\bibfield{author}{\bibinfo{person}{Kang Yang}, \bibinfo{person}{Miaomiao Liu}, {and} \bibinfo{person}{Wan Du}.} \bibinfo{year}{2024}\natexlab{}.
\newblock \showarticletitle{: Rateless-Enabled Link Adaptation for LoRa Networking}.
\newblock \bibinfo{journal}{\emph{IEEE/ACM Transactions on Networking}} (\bibinfo{year}{2024}).
\newblock


\bibitem[Zadrozny et~al\mbox{.}(2003)]%
        {zadrozny2003cost}
\bibfield{author}{\bibinfo{person}{Bianca Zadrozny}, \bibinfo{person}{John Langford}, {and} \bibinfo{person}{Naoki Abe}.} \bibinfo{year}{2003}\natexlab{}.
\newblock \showarticletitle{Cost-sensitive learning by cost-proportionate example weighting}. In \bibinfo{booktitle}{\emph{Third IEEE international conference on data mining}}. IEEE, \bibinfo{pages}{435--442}.
\newblock


\end{thebibliography}

\end{document}